\documentstyle[11pt,twoside,jltp,epsf]{article}
\title{On the Interplay of Disorder and Correlations}
\author{Karol I. Wysoki\'nski 
\address{Institute of Physics, M.
Curie-Sk\l{}odowska University,  ul. Radziszewskiego 10, 
Pl-20-031 Lublin, POLAND}}
\runninghead{K.I. Wysoki\'nski}{Interplay of Disorder and
Correlations} 
\begin{document} 
\begin{abstract}
I address here the question of the mutual interplay of strong
correlations and disorder in the system. I consider random version
of the Hubbard model. Diagonal randomness is introduced {\it via}
random on-site energies and  treated by the coherent potential
approximation. Strong, short ranged, electron - electron
interactions are described by the slave boson technique and found 
to induce additional disorder in the system. As an example I
calculate the  density of states of the random interacting binary
alloy and compare it with  that for non-interacting system.
\end{abstract}
\maketitle
\section{INTRODUCTION}
The list of materials in which carriers are strongly interacting
with each other and scatter by random impurities is quite long. It
comprises {\it  inter alia} high temperature superconducting
oxides \cite{HTS} and  various heavy  fermion alloys \cite{Kondo}.
Two models are usually used to study such system. It is either
Hubbard or Anderson model suitably extended to allow for the
description of real  materials. Here the single band Hubbard model
has been  used, as it is the simplest model of correlated system. 
The disorder is introduced into the model by allowing for
fluctuations of the local on-site energies $\varepsilon_{i}$. 
The main purpose of this work is to study the interplay of disorder
and correlations in the Hubbard model. I shall present analytical
and numerical calculations indicating that correlations induce
additional disorder in the system. {\it Mutatis mutandis} disorder
in the system affects the parameters of the Mott-Hubbard 
metal-insulator transition. In the recent studies of weakly
interacting disordered systems it has been found that due to
disorder the interactions scale to strong coupling limit \cite{DB}.
Here to treat many particle aspect of the problem we use the  slave
boson technique, which is known to be qualitatively valid for all 
strength of interaction. In particular this method reproduces the
results of  the Gutzwiller approximation at the saddle-point level
\cite{Br}. 

\section{THE MODEL AND APPROACH}

We start here with random version of the Hubbard model  

\begin{equation} 
H=\sum_{ij\sigma }t_{ij}c_{i\sigma }^{+}c_{j\sigma }+\sum_{i\sigma 
}(\varepsilon_i -\mu )c_{i\sigma }^{+}c_{i\sigma
}+U\sum_{i}n_{i\uparrow  }n_{i\downarrow }\,. 
\end{equation} 
The meaning of symbols is standard. The first term describes the hopping of 
carriers through the crystal specified by lattice sites $i$, $j$.  
$\varepsilon _{i}$ denotes the fluctuating site energy - it
introduces disorder into the model. $U$ is the repulsion  between two
opposite spin electrons occupying the same site.   
In the slave boson technique\cite{KotRuck} one introduces 4 auxiliary boson 
fields: $\hat e_i$, $\hat s_{i\sigma}$, $\hat d_i$ such that $\hat e^+_i\hat 
e_i$, $\hat s^+_{i\sigma}\hat s_{i\sigma}$, $\hat d^+_i\hat d_i$ project 
onto the empty, singly occupied by $\sigma$ and doubly occupied site $i$. 
The Hamiltonian (1) is in this enlarged Hilbert space  written
as   
\begin{equation} 
H = \sum_{ij\sigma} t_{ij} \hat z^+_{i\sigma}\hat z_{j\sigma} 
c^+_{i\sigma}c_{j\sigma} + U\sum_i \hat d^+_i\hat d_i +
\sum_{i\sigma}  (\varepsilon_i - \mu) c^+_{i\sigma}c_{i\sigma}\,, 
\end{equation} 
where $\hat z_{i\sigma} = (\hat 1 - \hat d^+_i\hat d_i - \hat s^+_{i\sigma} 
\hat s_{i\sigma})^{-1/2} (\hat e^+_i\hat s_{i\sigma} + \hat s^+_{i-\sigma} 
\hat d_i) \cdot (\hat 1 - \hat e^+_i\hat e_i - \hat s^+_{i-\sigma} \hat 
s_{i-\sigma})^{-1/2}$. Equation (2) is strictly equivalent to (1), when the 
constraints  
\begin{eqnarray} 
&& \hat e^+_i\hat e_i + \sum_\sigma \hat s^+_{i\sigma}\hat s_{i\sigma} + 
\hat d^+_i\hat d_i = 1  \nonumber \\ 
&& c^+_{i\sigma}c_{i\sigma} = \hat p^+_{i\sigma}\hat p_{i\sigma} + \hat 
d^+_i \hat d_i;~~~~~~\sigma = \uparrow, \downarrow 
\end{eqnarray} 
are fulfilled. 
In disordered system and at $T$ = 0 K, the mean field approach to slave 
bosons can be formulated by suitably generalizing the clean limit. One 
introduces the constraints into Hamiltonian with help of Langrange 
multipliers $\lambda _{i}^{(1)}$ and $\lambda _{i\sigma }^{(2)}$ and 
replaces boson operators $\hat{e}_{i}$, $\hat{s}_{i\sigma }$, $\hat{d}_{i}$ 
by classical, site dependent, amplitudes $e_{i}$, $s_{i\sigma }$, $d_{i}$. 
These amplitudes are calculated from the, configuration dependent, ground 
state energy $E_{GS}=\langle H\rangle $ by minimizing $E_{GS}$ with respect 
to all seven parameters $e_{i}$, $s_{i\sigma }$, $d_{i}$, $\lambda 
_{i}^{(1)} $, $\lambda _{i\sigma }^{(2)}$. As a result one gets three 
constraints: $e_{i}^{2}+\sum_{\sigma }s_{i\sigma }^{2}+d_{i}^{2}=1$, $ 
\langle c_{i\sigma }^{+}c_{i\sigma }\rangle =s_{i\sigma }^{2}+d_{i}^{2}$ and 
four additional equations which read  
\begin{eqnarray} 
&&\lambda _{i}^{(1)}e_{i}=-{\frac{1}{\xi _{i}}}{\frac{\partial \xi _{i}}{
\partial e_{i}}}\,{\rm Re}\left( \sum_{j\sigma }t_{ij}\xi _{i}\langle 
c_{i\sigma }^{+}c_{j\sigma }\rangle \xi _{i}\right)  \nonumber \\ 
&&(\lambda _{i}^{(1)}-\lambda _{i\sigma }^{(2)})s_{i\sigma }=-{\frac{1}{\xi 
_{i}}}{\frac{\partial \xi _{i}}{\partial s_{i\sigma }}}\,{\rm Re}\left( 
\sum_{j}t_{ij}\xi _{i}\langle c_{i\sigma }^{+}c_{j\sigma }\rangle \xi 
_{i}\right)  \nonumber \\ 
&&\left( U+\lambda _{i}^{(1)}-\sum_{\sigma }\lambda _{i\sigma }^{(2)}\right) 
d_{i}=-{\frac{1}{\xi _{i}}}{\frac{\partial \xi _{i}}{\partial d_{i}}}\left(  
{\rm Re}\sum_{j\sigma }t_{ij}\xi _{i}\langle c_{i\sigma }^{+}c_{j\sigma 
}\rangle \xi _{j}\right) 
\end{eqnarray} 
where $\xi _{i}=(1-d_{i}^{2}-s_{i\sigma }^{2})^{-1/2}(e_{i}s_{i\sigma 
}+s_{i-\sigma }d_{i})(1-e_{i}^{2}-s_{i-\sigma }^{2})^{-1/2}$. 
 
The important next step is connected with calculation of $\langle c_{i\sigma 
}^{+}c_{i\sigma }\rangle $ and $E_{i\sigma }=\sum_{j}t_{ij}\xi _{i}\langle 
c_{i\sigma }^{+}c_{j\sigma }\rangle \xi _{i}$ in the presence of disorder. 
In principle these quantities do depend on the particular distribution of 
all impurities (i.e. distribution of site energies $\varepsilon _{i}$). We 
shall calculate them from the corresponding Green's function. We use a 
version\cite{KIW} of the coherent potential approximation (CPA) to
calculate  (averaged and conditionally averaged) Green's functions.
To this end we  rewrite the mean field Hamiltonian in the form  
\begin{equation} 
\tilde{H}=\sum_{ij\sigma }t_{ij}\xi _{i}\xi _{j}c_{i\sigma }^{+}c_{j\sigma 
}+\sum_{i\sigma }(\varepsilon _{i}-\mu +\lambda _{i\sigma }^{(2)})c_{i\sigma 
}^{+}c_{i\sigma }+{\rm const}\,. 
\end{equation} 
Note, that Hamiltonian (5) contains not only the site dependent
parameters $ \varepsilon _{i}$ but due to correlations there appear
also $\lambda  _{i\sigma }^{(2)}$ which are the additional source
of, diagonal and spin  dependent, disorder. Besides that, the
parameters $\xi _{i}$, $\xi _{j}$  which vary from site to site make
effective hopings $\tilde{t}_{ij}=t_{ij}\xi _{i}\xi _{j}$ random
quantities. This off-diagonal (in  Wannier space) disorder is
particularly important. Fortunately  multiplicative dependence of
effective hopping on the kind of atoms at sites   $i$ and $j$ is
easy to handle within CPA.   

The procedure is standard\cite{KIW} and one gets the following equation for 
the density of states in paramagnetic phase  
\begin{equation} 
D(\varepsilon )=-{\frac{1}{\pi }}\,{\rm Im}\langle {\frac{\xi 
_{i}^{2}F[\Sigma (\varepsilon ^{+})]}{1-\left[ \Sigma (\varepsilon 
^{+})-(\varepsilon -\varepsilon _{i}+\mu -\lambda _{i}^{(2)})/\xi _{i}^{2}
\right] \,F[\Sigma (\varepsilon )]}}\rangle _{{\rm imp}} 
\end{equation} 
and the coherent potential $\Sigma (\varepsilon )$ is determined as a 
solution of equations  
\begin{eqnarray} 
&&\langle {\frac{\Sigma (z)-(z-\varepsilon _{i}+\mu -\lambda _{i}^{(2)})/\xi 
_{i}^{2}}{1-\left[ \Sigma (z)-(z-\varepsilon _{i}+\mu -\lambda 
_{i}^{(2)})/\xi _{i}^{2}\right] \,F[\Sigma (z)]}}\rangle _{{\rm
imp}}=0   \nonumber \\ 
&&F[\Sigma (z)]={\frac{1}{N}}\sum_{\vec{k}}{\frac{1}{\Sigma (z)-\varepsilon (
\vec{k})}}={\frac{1}{N}}\sum_{\vec{k}}\,\bar{G}_{\vec{k}}(z) 
\end{eqnarray} 
Here $\langle \cdots \rangle _{{\rm imp}}$ means averaging over disorder and  
$\varepsilon (\vec{k})={\frac{1}{N}}\sum_{ij}t_{ij}\,e^{-i\vec{k}(\vec{R} 
_{i}-\vec{R}_{j})}$ is electron spectrum in host (clean) material. 
 
To solve equations (4) one still has to calculate the quantities
$E_{i\sigma  }=\sum_{j}t_{ij}\xi _{i}\langle c_{i\sigma
}^{+}c_{j\sigma }\rangle \xi _{j}$  , which in general depend on
the configuration of all impurities in the  system. They can be
calculated from the knowledge of the CPA Green's functions in  the
following way. We assume that site $i$ is described by actual  
parameters {\it i.e.} $\varepsilon _{i}$, $\lambda _{i\sigma}^{(2)}$,
$\xi _{i}$, while  all other sites in the system are replaced by
effective ones described by  the coherent potential $\Sigma (z)$ and
the Green's functions $\bar{G}(z)$. Then  it is easy to find that
{\it e.g}.  
 \begin{equation}  E_{i\sigma }=\int d\omega
{\frac{1}{e^{\beta \omega }+1}}\sum_{j}t_{ij} \left(
-{\frac{1}{\pi }}\right) \,{\rm Im}\tilde{G}_{ij\sigma }(\omega
+i0)  \end{equation}  
\noindent and \cite{Economou}  
\begin{equation} 
\tilde{G}_{ij\sigma }(z)=\bar{G}_{ij\sigma }(z){\frac{1}{1-\tilde{\varepsilon 
}_{i\sigma }(z)\bar{G}_{ii\sigma }(z)}} 
\end{equation} 
Here $\tilde{\varepsilon}_{i\sigma }(z)=(z-\varepsilon _{i}+\mu -\lambda 
_{i\sigma }^{(2)})/\xi _{i}^{2}$, and by definition $\bar{G}_{ii\sigma 
}(z)\equiv F[\Sigma (z)]$ and $\bar{G}_{ij\sigma }(z)=\frac{1}{N}\sum_{\vec{k
}}\bar{G}_{\vec k}(z)e^{i\vec{k}(\vec{R}_{i}-\vec{R}_{j})}$ \ This
closes the  system of equations to be solved in a self-consistent
manner. In the next  section we show numerical calculations of the
effect of correlations on the  density of states in disordered
alloy.   

\section{NUMERICAL RESULTS AND DISCUSSION} 
 
For the purpose of numerical illustration of the general approach sketched 
in previous section we calculate here the density of states (DOS) of perfect 
interacting, random noninteracting and random interacting systems. For the 
purpose of numerical analysis we shall assume $U=\infty $ limit and {\it bcc} 
crystal structure, which leads to the following canonical
tight-binding spectrum  of noninteracting carriers in clean material 
 
\begin{equation} 
\varepsilon _{\vec{k}}=-8t\cos (k_{x}a)\cos (k_{y}a)\cos (k_{z}a). 
\end{equation} 
Using t=0.0625$eV$ leads to the noninteracting system bandwidth $W=1eV$. In 
the following all energies and frequencies are expressed in $eV$. The 
density of states corresponding to this spectrum is well known. It possesses 
Van Hove singularity in the middle of the band, which extends from $-8t$ to $ 
+8t$. The spectrum $\varepsilon _{\vec{k}}$ of a clean but interacting 
system in the $U=\infty $ limit is replaced by $\xi ^{2}\varepsilon _{\vec{k} 
}-\lambda ^{(2)}$. The corresponding density of states is thus of the same 
form but placed in the energy window $[-8t(1-n)-\lambda 
^{(2)},+8t(1-n)-\lambda ^{(2)}]$ and scaled by the factor $(1-n)^{-1}$. Note 
that in this $U=\infty $ limit the upper Hubbard band is pushed to infinity 
and not visible. 
 
\begin{figure}[tbp] 
\vspace{-1.0in}
{\epsfxsize=2.4in \epsfbox{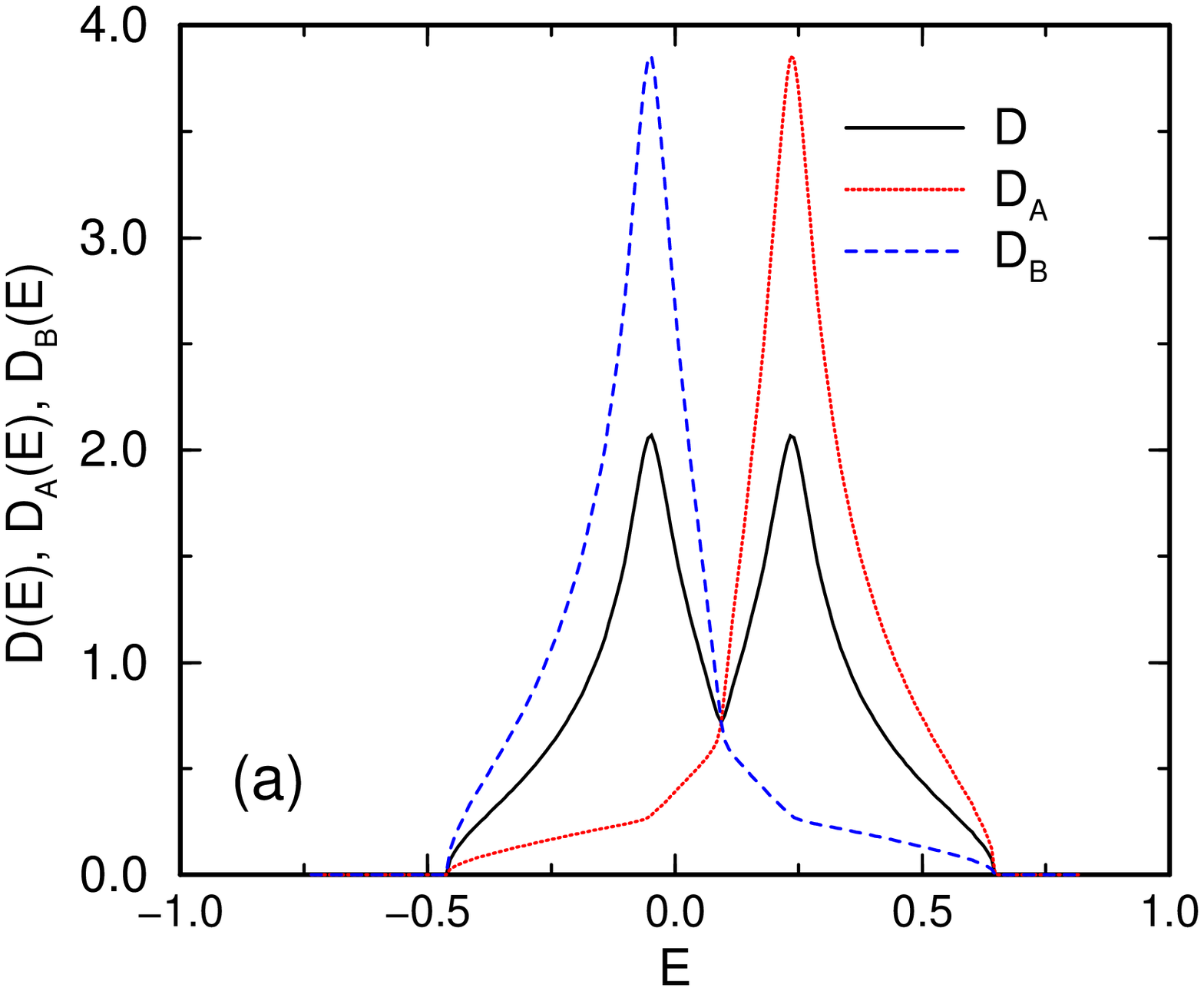}}
\hspace{0.12in}
{\epsfxsize=2.4in\epsfbox{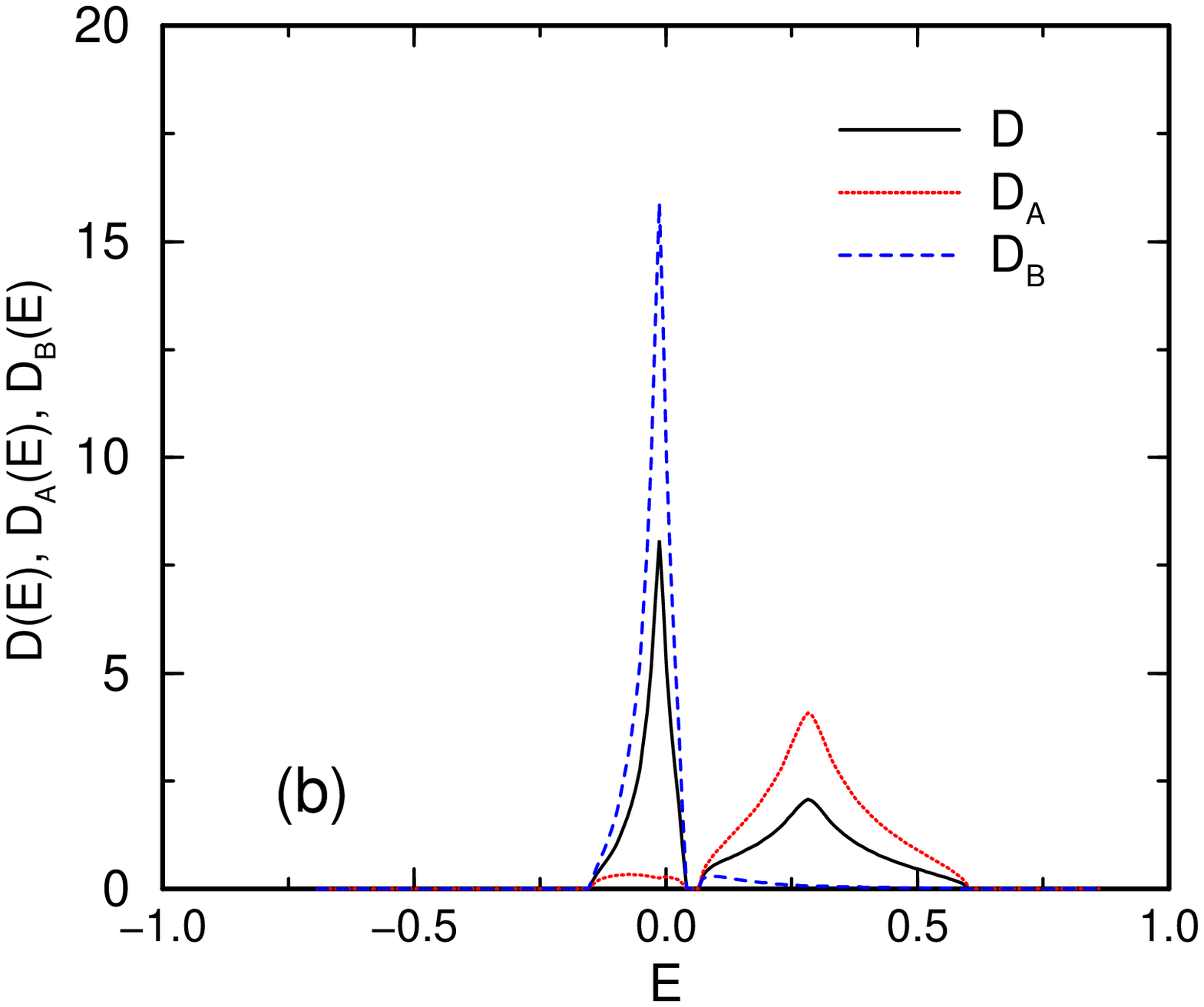}}
\caption{Comparison
of the DOS of (a) disordered noninteracting $ A_{1-x}B_{x} $ alloy
with $x=0.5$, $\protect\varepsilon _{A}=0$, $\protect \varepsilon
_{B}=-0.3eV$ with that calculated for the same alloy but with 
interacting carriers (b). Carrier density $n=0.4$. } 
\end{figure}

The changes of the spectrum due to correlations in disordered $A_{1-x}B_{x}$ 
alloy are illustrated in figure (1) where we show the averaged $D(E)$ and 
conditionally averaged densities of states $D_{A}(E)$ and $D_{B}(E)$ \ ($
x=0.5$, $\varepsilon _{A}=0,\varepsilon _{B}=-0.3eV$) without (Fig. 2a) and 
with (Fig. 2b) electron correlations. The carrier concentration $n=0.4$. 
Noninteracting alloy DOS is symmetric. We observe strong asymmetry, the 
opening of the real gap in the spectrum of interacting carriers and the 
appreciable increase of the density of states at the fermi level (taken as 
E=0 in the figure). 
 
In conclusion, we have shown that interplay of disorder and
correlations  leads to strong renormalisation of the electron
spectrum. Let us stress that  all other properties of such systems
will also be strongly affected by  interaction induced disorder. The
calculations of {\it dc} and {\it ac}  transport properties are in
progress.

\end{document}